# Teaching Programming to Novices: A Large-scale Analysis of App Inventor Projects

Nathalia da Cruz Alves
Department of Informatics and Statistics
Federal University of Santa Catarina
Florianópolis, Brazil
nathalia.alves@posgrad.ufsc.br

Christiane Gresse von Wangenheim
Department of Informatics and Statistics
Federal University of Santa Catarina
Florianópolis, Brazil
c.wangenheim@ufsc.br

Jean Carlo Rossa Hauck
Department of Informatics and Statistics
Federal University of Santa Catarina
Florianópolis, Brazil
jean.hauck@ufsc.br

*Abstract*—Teaching programming to K-12 students has become essential. In this context, App Inventor is a popular block-based programming environment used by a wide audience, from K-12 to higher education, including end-users to create mobile applications to support their primary job or hobbies. Although learning programming with App Inventor has been investigated, a question that remains is which programming concepts are typically used and how this compares to other block-based programming environments. Therefore, we explore the characteristics of App Inventor projects through a large-scale analysis of 88,606 apps from the App Inventor Gallery. We discovered that the size of App Inventor projects varies from projects with very few blocks to some surprisingly large projects with more than 60,000 blocks. In general, much fewer design components are used than programming blocks, as typically, to work properly, several programming blocks are necessary for each design component in an App Inventor project. In addition, we also compare our results with the analysis of 233,491 Scratch projects reported by Aivaloglou and Hermans [4]. Several differences can be observed, as in App Inventor projects events are more predominant, with lesser use of conditionals and loops. These findings may guide the decision on the adoption of App Inventor for teaching computing depending on the specific learning objectives or indicate the need for tailoring the curricula.

## I. INTRODUCTION

Nowadays, computing is evident in most activities. Regardless of the area of expertise of a professional, it is important not only to know the basic functionality and interfaces of computers but also the fundamentals and basic principles of computing. To teach computing programming concepts to novices and non-computing people, typically block-based visual programming environments are used such as App Inventor [1] or Scratch [2].

App Inventor is a visual programming environment that allows people to create mobile applications for Android devices through programming blocks rather than textual commands [1]. It is used by a wide range of people of all ages and backgrounds with more than 1 million unique monthly active users from 195 countries who created almost 35 million mobile apps [1]. App Inventor projects can be shared via the App Inventor Gallery [1] under the creative commons license, containing over 150,000 projects in April 2020. Another example is Scratch, a visual programming environment that allows people to create stories, animations, and games. It is designed especially for young people between 8 and 16 years, but can also be used by people of all ages [2]. Currently, it has more than 53 million users and more than 52 million projects shared within the Scratch community.

The evaluation of visual programming environments as tools for teaching computational thinking and end-user programming has received significant attention in the past. Several studies have been carried out on analyzing large datasets of projects created with visual programming environments, including Scratch [3][4][5][6][7][8], Snap! [9] and App Inventor [10][11]. These studies aim to identify the programming practices of programmer's projects in large galleries [3][4][9], learning trajectories [5], bad smells of programs [6], or skill progression [7][8][10][11].

There also exist studies comparing these visual programming environments regarding the effectiveness to learn computational thinking and introductory programming concepts [12][13][14]. Park and Shin [12] crawled mature projects, including tutorials and popular projects from the Scratch and App Inventor galleries. They used a rubric for comparing 524 Scratch mature projects with 379 App Inventor mature projects. Papadakis et al. [13][14] analyzed both visual programming environments regarding the pedagogical characteristics and their features [13][14]. They also performed an experiment with two experimental groups (Scratch and App Inventor) and one control group using Pascal with 87 Greek students to analyze the appropriateness of these programming environments to teach introductory programming in K-12 [14].

Yet, little is known about the programming practices of programmer's projects in large galleries, especially in App Inventor Gallery that includes projects of all types (not only mature projects). An analysis of App Inventor Classic has been done by Okerlund and Turbak [15], who extracted key features from 270,000 App Inventor projects created by 40,000 randomly chosen users. Yet, as this study refers to a previous version of App Inventor, their findings may be outdated. Xie et al. [16] analyzed a sample of 5,228 random projects grouping them by functionality to understand the usability and realized the capability of using App Inventor to implement specific functionalities. However, as the apps are grouped by their functionality, no detailed data on the frequency of the use of specific types of blocks is presented. In another study, Xie and Abelson [10] analyzed a sample of 10,571 random users, who each created at least 20 apps to analyze skill progression in App Inventor, focusing on a specific set of computational concepts blocks (CC-blocks) measuring learning progress. Thus, so far there is still missing a comprehensive overview of the current state of practice of App Inventor programs in large galleries, as well as how this compares to other popular block-based programming environments, such as Scratch.

Funding agencies: *Coordenação de Aperfeiçoamento de Pessoal de Nível Superior* - Brasil (CAPES) - Finance Code 001; *Conselho Nacional de Desenvolvimento Científico e Tecnológico* - Brasil (CNPq).

Therefore, the objective of this article is to obtain an understanding of how people program with App Inventor by analyzing apps shared via the public App Inventor Gallery [1]. We also aim at comparing the results with findings from the analysis of projects in a large gallery of another prominent visual programming language (Scratch) in order to identify similarities and differences, which may have implications on the instructional design of teaching programming depending on the adopted programming language.

This paper is organized as follows. Section II describes App Inventor programming environment. Section III introduces the methodology used in this research. Section IV and V presents the results and discussion on these results. Finally, Section VI presents our conclusions.

## II. APP INVENTOR

App Inventor is a visual open-source programming environment for Android devices used to create mobile applications [1]. It is a programming environment that uses a drag-and-drop blocks editor (Figure 1). It was originally provided by Google and it is currently run by the Massachusetts Institute of Technology. The current version 2.0 of App Inventor runs on a web browser, replacing App Inventor Classic.

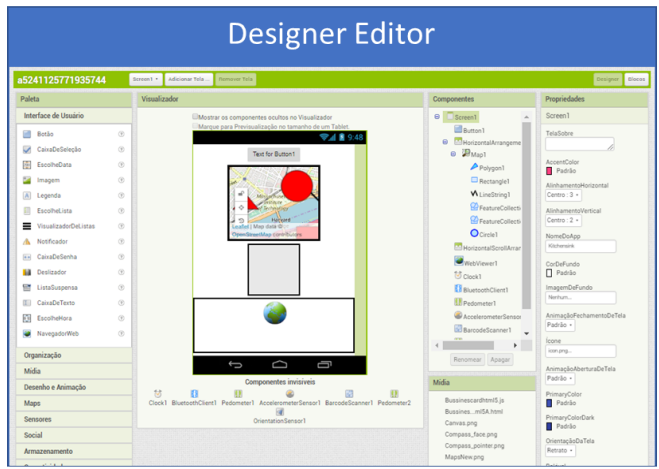

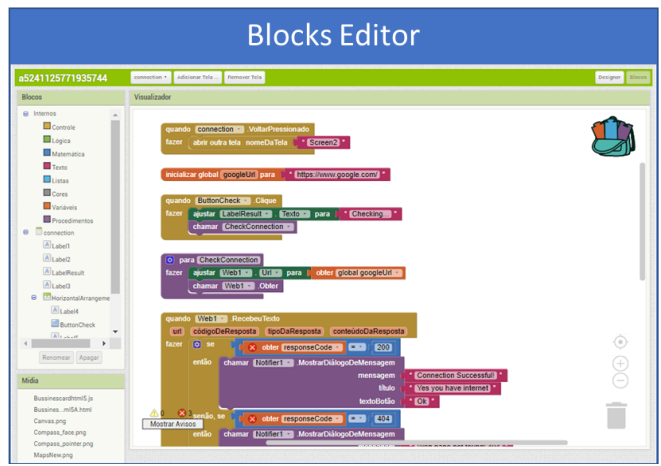

Fig. 1. Overview of the App Inventor Designer and Blocks Editor

App Inventor is taught and used by a wide audience, from K-12 to higher education, including end-user developers who write programs to support their primary job or hobbies [17][18].

A mobile app can be created in two phases with App Inventor. First, the Designer Editor configures the user interface components (Table 1), such as buttons, labels, etc. (Figure 1). The designer also allows non-visual components such as sensors, social, and media components that access phone features to be specified. App Inventor components are divided into categories (Table 1) with specific events, methods, and properties for each component.

TABLE I. OVERVIEW OF APP INVENTOR COMPONENTS

| Category | Description | Components examples |
|---|---|---|
| User Interface | Creating the visual part of the app. All visible components of the app are in this group. | Button, Checkbox, DatePicker, Image, Label, Notifier, etc. |
| Layout | Assists in organizing the visible components of the user interface category. | HorizontalArrangement, TableArrangement, etc. |
| Media | All media components from a device that can be used in apps. | Camcorder, Camera, Player, ImagePicker, Sound, etc. |
| Drawing and Animation | Components that allow the user to draw and view animations. | Ball, Canvas, ImageSprite |
| Maps | Maps components that include map navigation and markers. | Circle, Map, Marker, Polygon, etc. |
| Sensors | Components that get information from the sensors on the device. | AccelerometerSensor, Clock, GyroscopeSensor, etc. |
| Social | Components that allow the app to communicate with other social apps. | ContactPicker, EmailPicker, Sharing, etc. |
| Storage | Components that allow the creation of databases to store data. | File, FusiontablesControl, TinyDB, TinyWebDB |
| Connectivity | Components that allow application connectivity with other devices. | ActivityStarter, BluetoothClient, BluetoothServer, Web |
| Lego Mindstorms | Control of LEGO® MINDSTORMS® NXT robots using Bluetooth. | NxtDirectCommands, NxtColorSensor, NxtLightSensor, etc. |
| Experimental | Experimental components. | CloudDB, FirebaseDB |
| Extensions | Possibility of importing extensions with new blocks. | - |

The app's behavior is programmed in a second stage by connecting visual programming blocks in the Blocks Editor (Figure 1). Each block corresponds to abstract syntax tree nodes in traditional programming languages. Blocks can represent standard programming concepts like loops, procedures, conditionals, etc., or conditions, events, and actions for a particular component of the app or any component.

App Inventor has two main types of blocks: built-in blocks and component blocks. Built-in blocks are available for use in any app and refer to overall programming concepts, such as variables, conditionals, loops, procedures, logical and math operators, etc. Component blocks include events, set and get, call methods, and component object blocks that are available for specific design components added to the app (Table 2).

TABLE II.  OVERVIEW OF APP INVENTOR BLOCKS

| | Category (examples) | Description |
|---|---|---|
| Built-in blocks | Control | Blocks responsible for control commands including important blocks like loops, conditionals, and screen actions. Examples: controls_while, controls_if, controls_closeScreen. |
| | Logic | Blocks responsible for logic operations on variables including relational and Boolean. Examples: logic_compare, logic_operation. |
| | Math | Blocks responsible for creating numbers and perform basic and advanced math operations. Examples: math_add, math_cos. |
| | Text | Blocks responsible for creating and manipulating *original* strings. Examples: text, text_length. |
| | Lists | Blocks responsible for creating and manipulating *original* lists. Example: lists_create_with, lists_add_items. |
| | Colors | Blocks responsible for creating and manipulating colors. Examples: color_red, color_blue. |
| | Variables | Blocks responsible for creating and manipulating *original* variables. Examples: global_declaration, lexical_variable_set. |
| | Procedures | Blocks responsible for definition and call of *original* procedures. Examples: procedures_defnoreturn, procedures_callnoreturn. |
| Component blocks | Events | Blocks responsible for specifying how a component responds to certain events, such as a button has been pressed. Example: component_event |
| | Set and Get | Blocks responsible for change components properties. Example: component_set_get |
| | Call Methods | Blocks responsible for call component methods to perform complex tasks. Example: component_method |
| | Component object | Blocks responsible for getting the instance component. Example: component_component_block |

The source code files of an App Inventor project are automatically saved in the cloud, but they can also be exported as an .aia file. An .aia file is a compressed file collection that includes a project properties file, media files that the app uses, and two files are generated for each screen in the app: a .bky file and a .scm file. The BKY file wraps an XML structure including all the blocks of programming used to define the behavior of the app, while the .scm file wraps a JSON structure that contains all the used visual components in the app [19].

## III. DEFINITION AND DATA COLLECTION

The objective of this study is to understand the current state of practice of App Inventor projects in large galleries by analyzing its characteristics. Concerning this objective, we analyze the following questions:

AQ1. What is the size of App Inventor projects and what overall programming concepts (built-in blocks) are commonly used?

AQ2. Which programming blocks related to design components (component blocks) are commonly used in App Inventor projects?

AQ3. How do App Inventor projects differ from projects in Scratch?

With respect to these analysis questions, we defined metrics based on Aivaloglou and Hermans [4]. The size of a project is measured in the number of blocks. Similarly to Aivaloglou and Hermans [4], we infer through the use of a block in a project that the related programming abstraction or concept is being used. Question 2 is discussed based on an analysis of the use of components and its blocks. To answer which design components and component blocks are commonly used, we analyze all designer components and component blocks in App Inventor projects.

In order to answer analysis question 3, we compare our results on App Inventor projects with results reported by Aivaloglou and Hermans [4] being one of the most recent and largest analyses of 233,491 Scratch projects. We compare the mean with respect to characteristics of size, procedures, and programming concepts. The comparison of the usage of programming concepts is limited to concepts available in both programming environments, excluding, therefore, App Inventor designer components.

To answer these analysis questions, we conducted a quantitative analysis of data on App Inventor projects. With support from the MIT App Inventor team, we downloaded 88,863 App Inventor projects public available under the creative commons license in June 2018. Out of 88,863 available apps in the gallery, 256 projects failed to be analyzed due to technical difficulties. As a result, a total of 88,606 App Inventor projects has been analyzed.

We analyzed the App Inventor projects using an automated tool [20]. The tool is a web application that automatically assesses and grades projects programmed with App Inventor through static code analysis. The code analysis is done in three steps:

First, the project code (.aia file) is decompressed, read, parsed, and converted into a string to be manipulated more easily. Secondly, lexical analysis is performed on the resulting string, converting the sequence of characters into a sequence of tokens (strings with an assigned meaning). Finally, the tool goes through the token list, counting the frequency of each token, creating a table of tokens, and their frequency of use.

The analysis results were exported in a spreadsheet, which is available upon request from the first author. The spreadsheet includes the following data from the projects:

- Project ID
- Designer components per project
- Built-in blocks per project
- Component blocks per project

As App Inventor collects no demographic data on users such as gender, age, geographic location, programming background, etc., this type of information is not available. Through the App Inventor Gallery, there is also no information available on whether any of the projects were created individually or in collaboration with others, as part of a class/course or informally, etc.

## IV. RESULTS

In the following sections, we describe the results obtained through the analysis of the 88,606 App Inventor projects for each of the analysis questions.

*A. What is the size of App Inventor projects and what overall programming concepts (built-in blocks) are commonly used?*

To answer this question, we measure the size of projects based on the number of blocks and design components per project. The size of App Inventor projects varies from projects with very few blocks to some surprisingly large projects with more than 60,000 blocks. We found a mean of 162.5 blocks per project. However, a significant amount of App Inventor projects is rather small, in the first quartile 25% of the projects have up to only 15 programming blocks, whereas in the third quartile 75% of the projects have up to 137 blocks (Figure 2).

In general, much fewer design components are used than programming blocks (Figure 2), as typically to work properly, several programming blocks are necessary for each design component. We found a mean of 25.83 design components per project, and in the first quartile 25% of the projects have up to only 7 design components, whereas in the third quartile 75% of the projects have up to 24 design components (Figure 2).

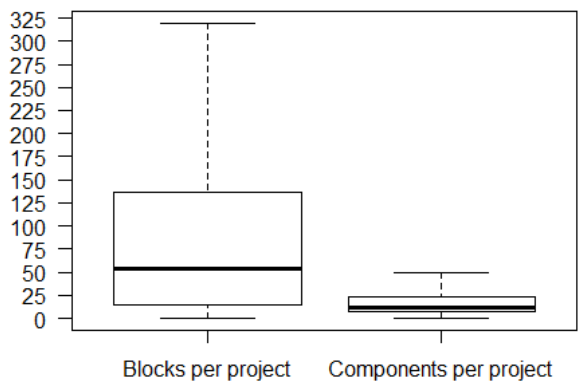

Fig. 2. Comparison of the number of programming blocks and design components per project (outliers removed)

Results show that 8% (7,022 out of 88,606 projects) of the apps seem to have no behavior, never change state, and/or enable user interaction as they have less than 2 programming blocks. As for an app to be interactive and have behavior, in addition to at least one design component, it must also have at least two programming blocks related to that component: One to handle an event and one to respond to that event [16].

Regarding programming concepts, we analyze statistics referring to the built-in blocks presented in Table 2, which includes concepts present in most programming languages such as variables, lists, logic, math operations, control blocks (loops and conditional), etc. In Table 3 we summarize the results presenting the mean value and five-number summary.

TABLE III. SUMMARY STATISTICS FOR PROGRAMMING BLOCKS

| Built-in blocks | mean | min | Q1 | median | Q3 | max |
|---|---|---|---|---|---|---|
| Control | 6.93 | 0 | 0 | 1 | 6 | 1943 |
| Logic | 8.97 | 0 | 0 | 0 | 4 | 7597 |
| Math | 19.94 | 0 | 0 | 1 | 14 | 7183 |
| Text | 19.58 | 0 | 0 | 3 | 14 | 7487 |
| Lists | 3.00 | 0 | 0 | 0 | 0 | 6514 |
| Colors | 1.93 | 0 | 0 | 0 | 0 | 4095 |
| Variables | 16.34 | 0 | 0 | 1 | 11 | 19690 |
| Procedures | 2.63 | 0 | 0 | 0 | 0 | 1454 |
| **Component blocks** | **mean** | **min** | **Q1** | **median** | **Q3** | **max** |
| Events | 9.08 | 0 | 2 | 5 | 10 | 886 |
| Set and Get | 23.98 | 0 | 1 | 6 | 19 | 10638 |
| Call methods | 7.07 | 0 | 0 | 2 | 6 | 3384 |
| Component object | 0.79 | 0 | 0 | 0 | 0 | 533 |

Among programming blocks, event blocks are by far the most used ones, being present in more than 91% of the projects (Figure 3). Of the 88,606 analyzed projects, 80,832 projects use at least 1 event command to handle interface events, timers, sensors, etc. This confirms the event-driven nature of the App Inventor programming language [21]. The other component blocks Set and Get, and Call Methods are also widely used to customize App Inventor components and perform complex functions that are encapsulated by the call methods.

Text blocks are used in 66% (58,430 out of 88,606 projects) (Figure 3) with a mean of 19.58 text blocks per project. Text blocks allow to create, manipulate text and are very useful for programming hardcoded functions.

Control blocks appear in more than 50% of the projects. However, we noticed that different control blocks (conditional, screen, and loop) have a different amount of use and in general, conditional blocks are much more used than loop blocks (Figure 4). Conditionals are used in 41% (36,649 out of 88,606 projects). These numbers are similar to the usage of logic blocks, as both concepts are typically used together to function properly. Loops are hardly used (Figure 4). With a mean of 0.23 loops per project, only 6% use loops

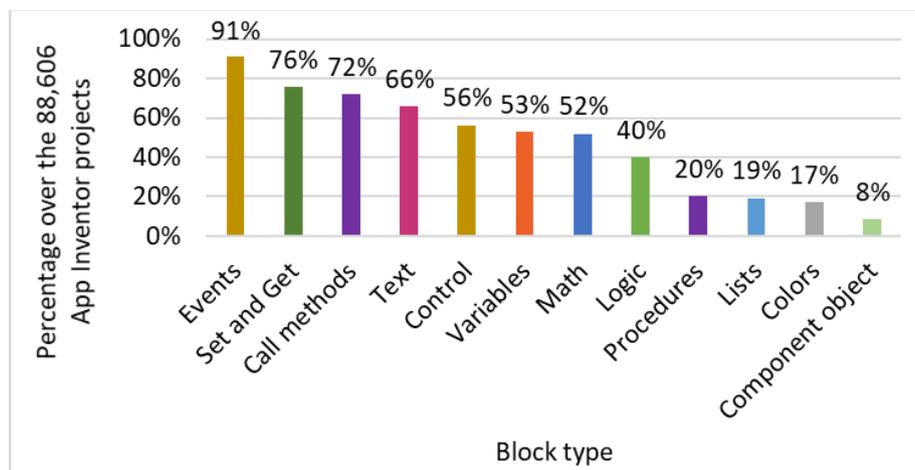

Fig. 3. Using component and built-in blocks

(5,635 out of 88,606 projects). This can be explained by the event-based model of App Inventor, in which many iterative processes, which would be expressed through loops in other languages, are expressed by an event that performs a single step of the iteration each time it is triggered [21].

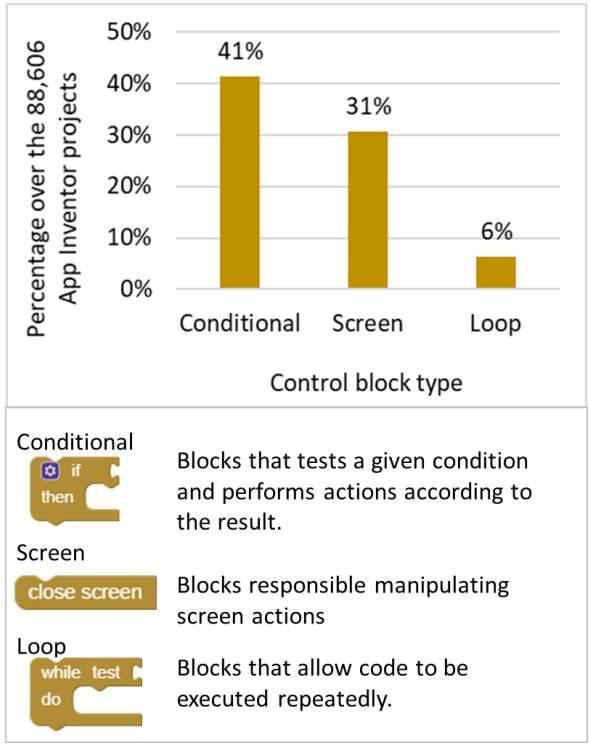

Fig. 4. Using control built-in blocks

Programming blocks related to variables are also among the most commonly used with a mean 16.34 blocks per project and 53% (46,959 out of 88,606 projects) use blocks related to creating and operating variables (Figure 3). However, it is important to note that the use of variable programming blocks per project (53%) is different from the creation of original variables per project (37%), which is related to creating variables only, in contrast to the use and operations of predefined variables. On the other hand, blocks related to lists, a more advanced structure for representing data, are used by only 19% (17,084 out of 88,606 projects) (Figure 3). Math blocks are used in 52% (45,902 out of 88,606 projects) (Figure 3) with a mean of 19.94 (Table 3). Some projects use several math blocks increasing the overall mean up to the maximum value of 7,183 math blocks in one project (Table 3).

Procedures blocks can be used both to abstract and to organize the code. However, only 20% of the projects use procedures blocks (Figure 3) with a mean of less than 3 procedure blocks per project (Table 3). To better understand the use of procedures, we also analyze procedures definition and procedures call for the set of projects with a least one procedure (17,895 of 88,606 projects) as shown in Table 4.

TABLE IV. SUMMARY STATISTICS FOR PROCEDURES PER PROJECT WITH PROCEDURES

| Procedures category | mean | min | Q1 | median | Q3 | max |
|---|---|---|---|---|---|---|
| Procedures *definition* per project with procedures | 3.62 | 1 | 1 | 2 | 4 | 286 |
| Procedure *call* per project with procedures | 9.38 | 0 | 2 | 4 | 9 | 1260 |

A higher mean (avg) of procedures calls than procedures definition in projects with a least one procedure indicates that procedures are being created not only for the organization but also for abstraction, as also pointed out by Xie and Abelson [10]. We also noticed that procedures without return values are defined and called over 7 times more often than procedures with return values (Figure 5). In the context of App Inventor, this indicates that procedures are often used to provide similar functionality to multiple components (e.g. 3 buttons providing color options for a painting app), rather than to perform repeated calculations with return values.

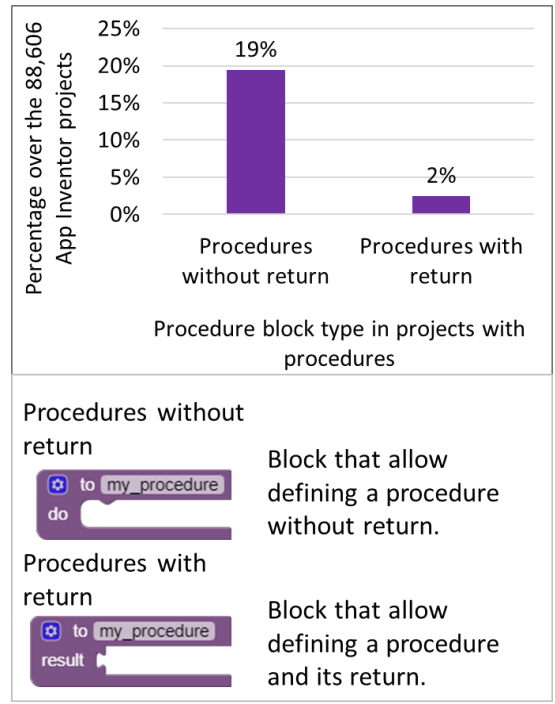

Fig. 5. Using procedures built-in blocks per projects with procedures

Color blocks seem to be the least explored programming built-in blocks as they are used in only 17% (15,311 out of 88,606 projects) (Figure 3) and only 0.2 million color blocks in general (Figure 6). This can be explained by the fact that these color blocks represent an advanced form for defining colors, which can be done much more easily through parameter settings in the Designer editor, without the need to use programming blocks.

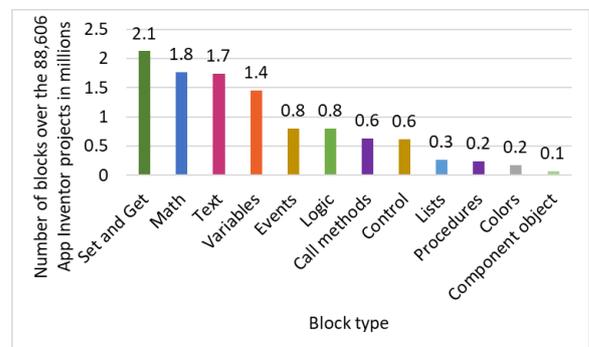

Fig. 6. Sum of using each blocks category

Overall, we observed differences between projects indicating a large number of rather simple programs and a few very complex ones. Although Events are used in the majority of App Inventor projects (Figure 3), the sum of event blocks in the analyzed projects is much smaller compared to Set and Get blocks (Figure 6).

The frequent use of set and get blocks can be explained by the need for using a lot of these blocks to change or know the values of many component properties, e.g. to set/get the text on-screen using a label. Math blocks are also used extensively, summing up 1.8 million math blocks in the analyzed projects (Figure 6). Component object blocks are used in only 8% of the 88,606 projects (Figure 3) and also rarely used when comparing their sum to other block categories (Figure 6), as they are typically used in more complex and advanced code structures that require a high level of abstraction.

## B. Which programming blocks related to design components (component blocks) are commonly used in App Inventor projects?

To answer this question, we analyze component blocks (events, set and get, call methods, and component object) for all design components (user interface, layout, media, drawing and animation, maps, sensors, social, storage, connectivity, Lego Mindstorms, and experimental). The descriptive statistics are shown in Table 5.

TABLE V. SUMMARY STATISTICS FOR COMPONENTS BLOCKS

| Component blocks | mean | min | Q1 | median | Q3 | max |
|---|---|---|---|---|---|---|
| User interface | 22.44 | 0 | 2 | 7 | 18 | 10,790 |
| Layout | 0.94 | 0 | 0 | 0 | 0 | 1,596 |
| Media | 2.73 | 0 | 0 | 0 | 3 | 1,378 |
| Drawing and Animation | 8.74 | 0 | 0 | 0 | 5 | 3,867 |
| Maps | 0.01 | 0 | 0 | 0 | 0 | 75 |
| Sensors | 1.65 | 0 | 0 | 0 | 1 | 1,306 |
| Social | 0.29 | 0 | 0 | 0 | 0 | 1,281 |
| Storage | 1.78 | 0 | 0 | 0 | 0 | 3,074 |
| Connectivity | 1.36 | 0 | 0 | 0 | 0 | 789 |
| Lego Mindstorms | 0.02 | 0 | 0 | 0 | 0 | 33 |
| Experimental | 0.04 | 0 | 0 | 0 | 0 | 169 |

We analyze the percentage of design components, as apps may contain a component, but do not necessarily have blocks related to that component for modifying its behavior (Figure 7).

In order to differentiate the use of design components vs. the use of component blocks, Figure 8 illustrates the use of the design component button vs. the use of the button component blocks (Figure 8).

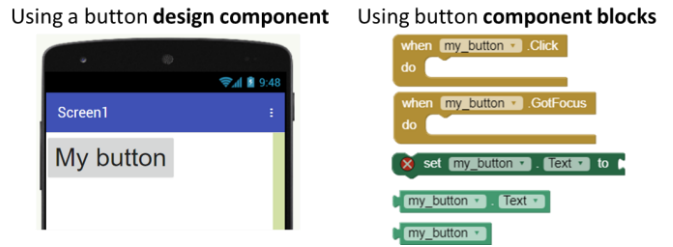

Fig. 8. Difference between using design components vs. using component blocks

The most commonly used component blocks are related to the user interface (UI) with a mean of 22.44 blocks per project (Table 5). As UI components are essential for creating the visual part of the app, this kind of component is also the only one that is present in all projects (93% of 88,606 projects) (Figure 7). The component blocks related to interface design components are also the most frequently used as 86% of projects contain user interface blocks to change its behavior for the interaction with the user.

Many projects also use layout components to organize the visible elements of the user interface (61% out of 88,606) (Figure 7). However, layout blocks are rarely used in only 5.4% of projects. This may indicate a stronger focus on the use of these components to statically organize the screen interface rather than dynamically organize the user interaction, e.g., changing the local of a button.

Other components interacting with features of mobile devices such as Media components (including camera, sound recorder, speech recognizer, player, etc.), Sensors (including accelerometer, gyroscope, location, etc.) and Social (including contact picker, e-mail picker, phone call, etc.) are used in some projects (Figure 7), as they are applicable only for specific types of applications not representing a general type of component required for any kind of app.

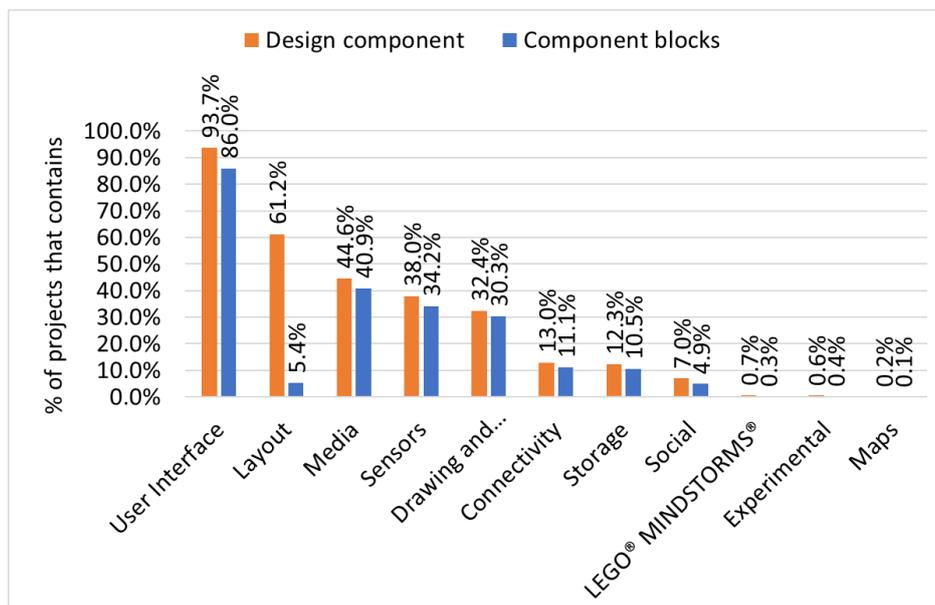

Fig. 7. Using design components vs. using component blocks

Less than 12% of the projects use Connectivity blocks (Figure 7) to open other applications of the phone or transfer data through Bluetooth and/or the Internet. Map blocks are rarely used by only 0,1% (177 out of 88,606 projects). Yet, this component may be underrepresented in our dataset as it has been added only recently to the App Inventor environment and, thus, older projects do not use Maps. Lego Mindstorms components enabling physical computing by apps interacting with Lego constructions are hardly used by only 0,7% (664 out of 88,606 projects), as they represent a very specific type of app.

Analyzing specifically design component and component blocks for the user interface category, the most commonly used visible user interface elements are buttons (85% of 88,606 projects), followed by labels (63% of 88,606 projects) and text boxes (31% of 88,606 projects) (Figure 9). Among the less commonly used user interface components are the time picker (1% of 88,606 projects) and date picker (2% of 88,606 projects).

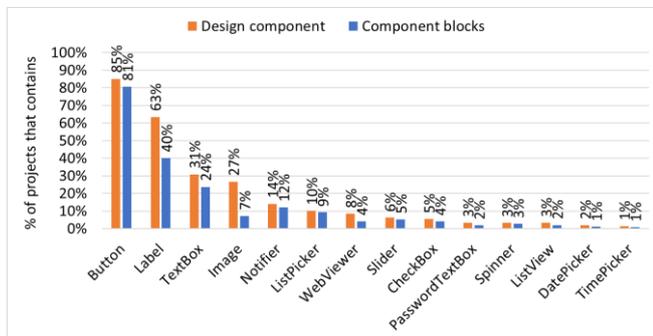

Fig. 9.   Using UI design components vs. using UI components blocks

We also observed that storage design components (File, FusiontablesControl, TinyDB, and TinyWebDB) are used by only 12.3% (10,887 out of 88,606 projects) (Figure 7), and experimental design components, related to database components (CloudDB and FirebaseDB) are rarely used by only 0,6% (542 out of 88,606 projects). The most used design component for data persistence is TinyDB by 10% (8,875 out of 88,606 projects). The others are scarcely used by less than 2% of projects in general, with the CloudDB being the least used component for data persistence by only 47 projects (Figure 10).

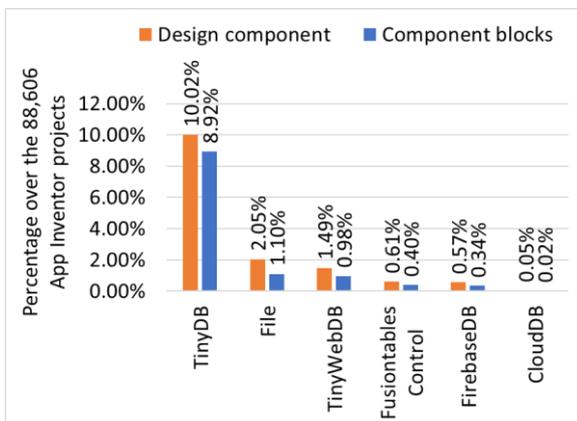

Fig. 10. Using Storage and Experimental design components vs. using Storage and Experimental component blocks

A considerable number of projects from the App Inventor Gallery also use components related to Drawing and Animation, in which the Canvas component, which is more related to the layout, is used in 32% of the projects. However, Canvas blocks are used in only 22% of projects (Figure 11). In comparison ImageSprint and Ball, Drawing and Animation components more related to animation, have similar use in components and blocks (Figure 11). Considering that several App Inventor beginner tutorials [22] involve the creation of a drawing app, this usage of Drawing and Animation blocks may be due to the fact that many users create projects that are very similar in functionality to these tutorials as also pointed out by Xie et al. [16].

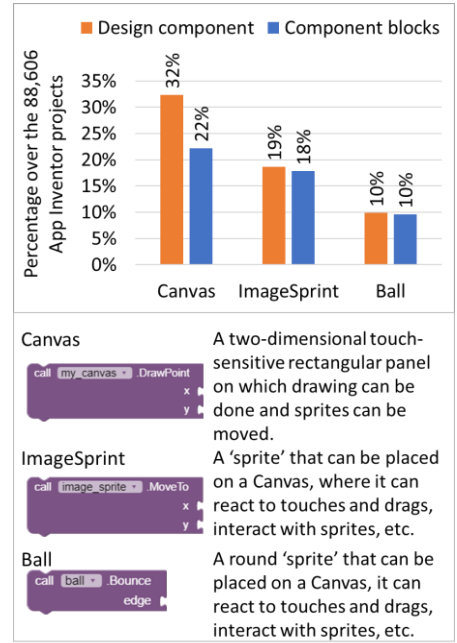

Fig. 11. Using Drawing and Animation design components vs. using Drawing and Animation component blocks

### C. How do App Inventor projects differ from projects in Scratch?

To answer this question, we compare our results of 88,806 App Inventor projects with the results reported by Aivaloglou and Hermans [4] analyzing 233,491 Scratch projects, being Scratch one of the most prominent visual programming languages. The comparison is limited to concepts available in both programming environments, excluding, therefore, App Inventor designer components. The descriptive statistics are shown in Table 6.

TABLE VI.   SUMMARY STATISTICS OF THE DATASET OF 88,606 APPINVENTOR PROJECTS COMPARED WITH THE DATASET OF 233,491 SCRATCH PROJECTS FROM AIVALOGLOU AND HERMANS [4]

| Item description | Language | mean | min | Q1 | median | Q3 | max |
|---|---|---|---|---|---|---|---|
| Number of blocks per project | Scratch | 154.55 | 1 | 12 | 29 | 76 | 34,622 |
| | AppInventor | 162.50 | 0 | 15 | 54 | 137 | 60,709 |
| Variables per project | Scratch | 2.06 | 0 | 0 | 0 | 1 | 340 |
| | AppInventor | 1.93 | 0 | 0 | 0 | 2 | 1673 |
| Lists per project | Scratch | 0.55 | 0 | 0 | 0 | 0 | 317 |
| | AppInventor | 1.07 | 0 | 0 | 0 | 0 | 1938 |
| Conditional blocks per project | Scratch | 10.02 | 0 | 0 | 0 | 3 | 5950 |
| | AppInventor | 4.03 | 0 | 0 | 0 | 2 | 1457 |
| Loop blocks per project | Scratch | 7.65 | 0 | 1 | 2 | 5 | 2503 |
| | AppInventor | 0.23 | 0 | 0 | 0 | 0 | 608 |
| Procedures per project with procedures | Scratch | 11.50 | 1 | 1 | 2 | 6 | 847 |
| | AppInventor | 3.62 | 1 | 1 | 2 | 4 | 286 |
| Calls per procedure | Scratch | 2.14 | 0 | 1 | 1 | 2 | 526 |
| | AppInventor | 9.38 | 0 | 2 | 4 | 9 | 1260 |

In general, App Inventor projects seem to be larger in terms of the number of blocks per project than Scratch projects (Table 6). Although the mean number of blocks per project (162.50 blocks in App Inventor projects and 154.55 blocks in Scratch projects) do not differ much, the median of blocks per project in App Inventor (54) is almost twice the median of blocks per Scratch project (29). Therefore, it seems that App Inventor projects, in general, seem to be more complex than programs created with Scratch.

The median number of variables per project is similar for both languages. On the other hand, lists are more frequently used in App Inventor than Scratch projects (Table 6), with almost twice a mean of lists per project.

Other significant differences can be observed concerning the usage of conditional blocks. App Inventor projects present a much lower mean of conditional statements per project than Scratch projects (Table 6). A possible explanation could be the fact that in App Inventor projects some conditional behaviors are expressed implicitly through event blocks (e.g. "when a button is clicked, do something").

Another expressive difference concerns loops, as the mean of loops per App Inventor project (0.23) is significantly lower than in Scratch projects (7.65). This shows that due to the event-based model of App Inventor many iterative processes that are expressed through loops in Scratch are represented using event blocks in App Inventor. In addition, the limitation of the use of App Inventor threads [21] discourages frequent use of loops, as they may crash the application.

Projects in Scratch also include more procedure definitions than App Inventor projects (Table 6). However, when procedures are used, projects in App Inventor call the defined procedures more often than Scratch projects, in which procedures typically are only called once or twice from the same script. This seems to indicate that in Scratch procedures are more used for organization, whereas procedures in App Inventor are used for the organization as well as abstraction.

## V. DISCUSSION

Based on the analysis of 88,606 apps from the App Inventor Gallery, we can observe a large variety of apps ranging from very small (even nonfunctional apps) to quite large apps with more than 60,000 blocks. The usage of programming blocks centers clearly on the usage of component blocks in which events blocks are used in more than 90% of the projects due to the event-based nature of App Inventor. Other kinds of blocks such as conditionals and loops are less used, especially when compared to other languages, as their function is partially taken over by events. Programming blocks related to variables are also among the most commonly used, while blocks related to lists are less used.

Concerning designer components, several user interface components seem to be essential to apps as they are present in the majority of the apps, such as buttons, labels, and text boxes. Other user interface components, such as date picker, time picker, and spinner are more specific required only by certain types of apps. Blocks related to layout and drawing and animation (canvas) are also relatively frequently used indicating their importance for the organization of user interface elements. Other categories of designer components are used more sparsely, due to their specific objective. Not all apps require the access to sensors of the mobile device (such as GPS, accelerometer, etc.), media (such as camera, speech recognition, etc.), social (such as e-mail picker, contact picker, phone call, etc.) or connectivity via Bluetooth, Internet, etc. Surprisingly few apps use storage blocks, mostly only TinyDB, as a simple database in order to store app data permanently.

These results of our analysis are also consistent with the results presented by Xie and Abelson [10]. Our analysis also indicates that among the CC-blocks (Computational Concept), which do not include text blocks (the most used built-in blocks), variable and control are the most used blocks in projects. The least used CC-blocks, which do not include color blocks (the least used built-in blocks), are list commands. However, different from the results presented by Xie and Abelson [10], we found that procedures are called more often than defined.

Based on the results of the analysis of the state of the practice of App Inventor projects, especially when compared to other visual programming languages such as Scratch, we can identify several implications for teaching computing. Adopting App Inventor, students are introduced to an event-based processing model building mobile apps, which makes some concepts easier to learn than others, such as component properties and event parameters as also pointed out by Xie [11]. On the other hand, other concepts like conditionals and/or loops are less required and, therefore, much less covered when teaching computing using App Inventor.

Concepts such as lists, which are used only by few App Inventor projects, yet almost twice as often as in Scratch projects, seem to represent more complex concepts used by more experienced users. This may also apply to other concepts such as procedures and storage. Xie et al. [16] also comment that the storage group seems to represent an advanced concept, as storage components often require structures such as lists and loops to leverage its functionality. These results may indicate that as part of a curriculum, these concepts should be approached in advanced courses/tutorials rather than beginner ones, for example, placing easier concepts first and more difficulty ones later [24]. Another aspect that gains importance in teaching computing by building mobile apps, is the user interface design as a key factor for the success of an app. Thus, teaching computing in this context should also cover basic competencies on user interface design, including hierarchy, color, typography, and imagery.

The results of this analysis of the state of the practice of App Inventor projects can help to develop computing curriculums tailored to the specific characteristics of App Inventor taking benefit of its events-based model. They also may guide the decision of adopting App Inventor or a different visual programming language depending on the learning objectives in a specific context. For example, when aiming at teaching conditionals and/or loop concept, the usage of Scratch may be more indicated. Identifying more advanced components in App Inventor may also guide the allocation of certain content to different levels of learners.

**Threats to validity**. This work is subject to various threats to validity. To minimize their impact in our research we identified potential threats and applied mitigation strategies.

In order to mitigate threats related to the design of the study, we defined and documented a systematic methodology for our study using the GQM approach [23]. Concerning risks related to validity, we scraped a random sample from the App Inventor Gallery analyzing 88,606 projects downloaded in June 2018 that were automatically analyzed using an automated tool [20]. Thus, features more recently added to App Inventor, such as maps, may not have been covered by this sample to the same extent as features that have been available for a longer time. Furthermore, we use the number of blocks in the projects as a measure for the length of a program. Although this does not exactly correspond to the "length" of a project in lines, we assume that the number of blocks is an adequate measure for size, also used in other similar researches [4][11].

Concerning external validity, we used data collected from the App Inventor Gallery, the main public place for publishing and sharing App Inventor projects. In terms of statistical significance, a sample of more than 88,000 apps is a satisfactory sample size allowing the generation of significant results. Our comparison is based on publications with 233,491 Scratch projects [4] and at least 20 App Inventor projects each of 10,571 random users [11] also representing significant sample sizes.

To mitigate threats in terms of reliability, we documented a systematic methodology, defining clearly the study objective, the process of data collection, and the statistical methods used for data analysis. Furthermore, the research has been conducted by researchers with a background in computing and statistics.

## VI. CONCLUSIONS

In this paper, we present the results of a large-scale study on 88,606 projects we scraped from the App Inventor Gallery. We analyze these projects in terms of size, utilization of programming built-in blocks, design components, and components blocks. We also compare App Inventor projects to projects in Scratch as another prominent visual programming language. Our findings demonstrate the expressive use of event blocks due to the event-based model of App Inventor, whereas other computing concepts, such as conditionals and loops are not frequently used. Concepts, including media, sensors, social, and connectivity are less used due to their need only in specific kinds of apps. Concepts like lists, procedures, and storage seem to represent more advanced concepts, requiring more complex structures to leverage their functionality.

This analysis of the state of the practice of App Inventor projects may guide decisions on the selection of a visual programming language for teaching computing in a more systematic way as well as support the development of curriculums tailored to the specific characteristics of App Inventor. Our analysis also points out the importance of teaching user interface design as part of computing education and programming when using App Inventor to build mobile apps.

ACKNOWLEDGMENT

We would like to thank all researchers from the MIT App Inventor team, who provided support for the access to the App Inventor Gallery. The authors would also like to thank the anonymous referees for their valuable comments and helpful suggestions.